\documentclass[apl,preprint,onecolumn,superscriptaddress]{revtex4}

\usepackage{latexsym}
\usepackage{graphicx}
\usepackage{braket}

\begin{document}

\author{Z. Kim}
\affiliation{Joint Quantum Institute and Department of Physics, University of
Maryland, College Park, Maryland, 20742, USA}
\author{C. P. Vlahacos}
\affiliation{Center for Nanophysics and Advanced Materials, College Park, Maryland, 20742, USA}
\author{J. E. Hoffman}
\affiliation{Joint Quantum Institute and Department of Physics, University of
Maryland, College Park, Maryland, 20742, USA}
\author{J. A. Grover}
\affiliation{Joint Quantum Institute and Department of Physics, University of
Maryland, College Park, Maryland, 20742, USA}
\author{K. D. Voigt}
\affiliation{Joint Quantum Institute and Department of Physics, University of
Maryland, College Park, Maryland, 20742, USA}
\author{B. K. Cooper}
\affiliation{Joint Quantum Institute and Department of Physics, University of
Maryland, College Park, Maryland, 20742, USA}
\affiliation{Center for Nanophysics and Advanced Materials, College Park, Maryland, 20742, USA}
\author{C. J. Ballard}
\affiliation{Trevecca Nazarene College, Nashville, Tennessee, 37210, USA}
\author{B. S. Palmer}
\affiliation{Laboratory for Physical Sciences, College Park, Maryland, 20740, USA}
\author{M. Hafezi}
\affiliation{Joint Quantum Institute and Department of Physics, University of
Maryland, College Park, Maryland, 20742, USA}
\affiliation{National Institute of Standards and Technology, Gaithersburg, Maryland, 20899, USA}
\author{J. M. Taylor}
\affiliation{Joint Quantum Institute and Department of Physics, University of
Maryland, College Park, Maryland, 20742, USA}
\affiliation{National Institute of Standards and Technology, Gaithersburg, Maryland, 20899, USA}
\author{J. R. Anderson}
\affiliation{Joint Quantum Institute and Department of Physics, University of
Maryland, College Park, Maryland, 20742, USA}
\affiliation{Center for Nanophysics and Advanced Materials, College Park, Maryland, 20742, USA}
\author{A. J. Dragt}
\affiliation{Joint Quantum Institute and Department of Physics, University of
Maryland, College Park, Maryland, 20742, USA}
\author{C. J. Lobb}
\affiliation{Joint Quantum Institute and Department of Physics, University of
Maryland, College Park, Maryland, 20742, USA}
\affiliation{Center for Nanophysics and Advanced Materials, College Park, Maryland, 20742, USA}
\author{L. A. Orozco}
\affiliation{Joint Quantum Institute and Department of Physics, University of
Maryland, College Park, Maryland, 20742, USA}
\author{S. L. Rolston}
\affiliation{Joint Quantum Institute and Department of Physics, University of
Maryland, College Park, Maryland, 20742, USA}
\author{F. C. Wellstood}
\affiliation{Joint Quantum Institute and Department of Physics, University of
Maryland, College Park, Maryland, 20742, USA}
\affiliation{Center for Nanophysics and Advanced Materials, College Park, Maryland, 20742, USA}

\title{Thin-film superconducting resonator tunable to the ground-state hyperfine splitting of $^{87}$Rb}

\newcommand{\etal}{\textit{et al.}}
\newcommand{\kb}{k_{B}}

\date{\today}

\begin{abstract}
We describe a thin-film superconducting Nb microwave resonator, tunable to within 0.3 ppm of the hyperfine splitting of $^{87}$Rb at $f_{Rb}=6.834683$ GHz.
We coarsely tuned the resonator using electron-beam lithography, decreasing the resonance frequency from 6.8637 GHz to 6.8278 GHz.
For \emph{in situ} fine tuning at 15 mK, the resonator inductance was varied using a piezoelectric stage to move a superconducting pin above the resonator.
We found a maximum frequency shift of about 8.7 kHz per 60-nm piezoelectric step and a tuning range of 18 MHz.

\end{abstract}

%\pacs{84.40.Dc}
\maketitle

%\section{Introduction}

A hybrid quantum system composed of neutral atoms coupled to superconducting qubits~\cite{Wallquist2009,Andre2006} offers potential advantages as a platform for quantum computing, including high processing speed (10 ns) with a superconducting component and long-lived (seconds) storage in an atom-based qubit~\cite{Martin2006}.
However, significant practical difficulties are apparent in realizing such a hybrid system due to the very weak coupling between the two components and the very different requirements for manipulating neutral atoms and operating superconducting devices.
The recent realization of one-dimensional arrays of trapped neutral atoms around sub-$\mu$m diameter optical fibers~\cite{Vetsch2010} represents a significant step towards overcoming some of these difficulties.
In particular, no magnetic field is required for trapping (minimizing interference with superconducting devices) and relatively little optical power escapes the fiber~\cite{Vetsch2010} (reducing heating).
In addition, the atoms are trapped tightly along a straight line defined by the fiber, enabling thousands of atoms to be coupled uniformly to the magnetic field produced in a sub-mm straight section of wire in a superconducting circuit.

In this Letter, we report on a thin-film superconducting resonator that could be used as a tunable coupling element in an atom-superconductor hybrid system~\cite{Imamoglu2009}.
We demonstrate coarse and fine-frequency tuning of the resonator through the $^{87}$Rb hyperfine transition while maintaining a high quality factor.
For coarse tuning, we used e-beam lithography to add a small capacitance to the resonator and decrease the resonance frequency.
For fine tuning, we employed a movable superconducting pin to change the inductance of the resonator.

A $^{87}$Rb atom has hyperfine ground states, $|^{5}S_{1/2};F=1\rangle$ and $|^{5}S_{1/2};F=2\rangle$, which in zero magnetic field are separated by $f_{Rb}=$ 6.834683 GHz. This frequency is readily accessible to thin-film superconducting resonators and qubits.
The natural coupling between $^{87}$Rb atoms and a superconducting resonator is the magnetic dipole of the atom interacting with the magnetic field produced by current in the resonator's inductor.
The expected coupling strength between a few thousand atoms in a 1D array and a small resonator is a few kHz~\cite{Hoffman2011}.
For such weak coupling to be measurable, one needs a stable but precisely tunable resonator that has a high quality factor Q~\cite{Creedon2011,PaikBAPS2011}.
Such a resonator would also be useful to study charged two-level defects in dielectrics and solid state qubits~\cite{MartinisPRL2005,GaoAPL2008,ZaeillPRB2008,ZaeillPRL2011}.

We designed and fabricated a high-$Q$ superconducting resonator from thin-film Nb.
The resonator consists of a meander-line inductor and interdigitated capacitor coupled to a
transmission line (see Fig.~\ref{fig:Fig1}).
A 300 nm thick layer of Nb was sputtered on a c-plane sapphire wafer and then patterned using standard photolithography and an SF$_{6}+\rm{O}_{2}$ plasma etch.
Microwave simulations of the quasilumped-element resonator showed no other modes up to 28 GHz~\cite{ZaeillPRL2011}.

The resonator was mounted on a printed circuit board, packaged in a Cu box, and bolted to an rf shield, which was attached to the mixing chamber of a Triton 200 Cryofree dilution refrigerator (see Fig.~\ref{fig:Fig2}).
For \emph{in situ} fine tuning, we positioned a 99.9999\% pure Al pin above the resonator.
The Al pin was secured to a piezoelectric translation stage (ANPz101, attocube systems), which was grounded via a thin Cu foil to the rf shield.
The pin has a diameter of 625 $\mu m$ and overlaps the entire resonator.
The pin is also grounded to the rf shield through metalized fabric to reduce coupling of external noise to the resonator.

We find the resonance frequency by applying microwaves to the transmission line and measuring the output power $P_{out}$ while sweeping the source frequency.
To reduce Johnson-Nyquist noise from higher temperatures, the input microwave line has 10 dB of attenuation at 3.6 K, 10 dB at 0.7 K, and 30 dB at 15 mK [see Fig.~\ref{fig:Fig2}(b)].
On the output line, two isolators are mounted on the mixing chamber and each has a minimum isolation of 18 dB between 4 and 8 GHz.
The output microwave signal is amplified with a low noise amplifier sitting on the 3.6 K plate.
We used a vector network analyzer (VNA, Agilent E5071C) to generate the input microwave signal and detect the transmitted output with the VNA locked to an atomic clock (FS725, Stanford Research Systems).
At 15 mK, the measured resonance frequency of the resonator was 6.8637 GHz, about 29 MHz above $f_{Rb}$.
Given the sensitivity of the resonance frequency to the precise dimension of the resonator, this 29 MHz difference is attributable to normal processing variations and the limited precision of the design simulation.

Instead of building a new resonator, we used e-beam lithography to increase the capacitance of the resonator and thereby adjust the resonance frequency to slightly below $f_{Rb}$.
This approach allows us to make a small precise change in the resonance frequency of the resonator.
Simulations using Microwave Office (AWR Corporation) show that the resonance frequency decreases by about 80 MHz when a 100 $\mu m$ long finger is added to the interdigitated capacitor or by about 0.8 MHz for a 1 $\mu m$ increase in the length of a finger [see Fig.~\ref{fig:Fig2}(c)].
We used a scanning electron microscope to pattern a bilayer of MMA(8.5)MMA copolymer and ZEP520A and then deposited an Al pad (20 $\mu$m $\times$ 5 $\mu$m $\times$ 200 nm thick) on one finger of the Nb interdigitated capacitor [see Fig.~\ref{fig:Fig1}(b)].

With this small change to the capacitor, the measured resonance frequency decreased by 35.9 MHz to 6.8278 GHz, which is 1 part in 1,000 below $f_{Rb}$ [see Fig.~\ref{fig:Fig3}(a)].
Figure~\ref{fig:Fig3}(b) shows that the long-term drift of the resonance frequency is less than 1 kHz over 70 hours, a stability of better than 2 ppb/hr, possibly limited by our measurement system.

Figure~\ref{fig:Fig4}(a) shows the measured resonance frequency $f_{r}$ of the resonator around $f_{Rb}$ versus the separation between the Al pin and the resonator (red circles).
The pin was initially positioned about 600 $\mu m$ above the resonator, and, as it approached the resonator, the resonance frequency increased nonlinearly by over 18 MHz.
This nonlinear tuning is predominantly due to the pin screening the self-inductance $L$ of the resonator~\cite{inductance},
\begin{equation}
L\simeq L_{0}(1-M^2/L_{0}^2),
\label{eq:inductance}
\end{equation}
where $L_{0}$ is the un-screened self-inductance of the resonator and $M$ is the mutual inductance between $L$ and the image currents produced in the pin.
The resonance frequency is then $f_{r}=1/2\pi\sqrt{L C}$.
The complex geometry of the system requires 3D microwave simulations to accurately model the nonlinear tuning.
We calibrated the distance between the Al pin and the resonator by comparing the measured resonance frequency with the simulations [blue squares in Fig.~\ref{fig:Fig4}(a)] and found that the closest height of the pin was about 40 $\mu m$ at a tuned frequency of $f_{r}=6.8454$ GHz.

Figure~\ref{fig:Fig4}(b) shows the dependence of the tuning resolution on the step voltage applied to drive the piezo stage.
For a  36 $V$ step voltage, the frequency change per 60-nm step was about 8.7 kHz [see Fig.~\ref{fig:Fig4} (c)].
Figure~\ref{fig:Fig4}(d) shows a detailed view of the 30 $V$ data (blue curve) in Fig.~\ref{fig:Fig4}(b) around $f_{Rb}$.
Each point was measured after 8 steps of the piezo stage.
Here the tuning resolution was a few kHz for 8 steps (1 ppm) or less than 1 kHz per step on average (100 ppb).
Below 30 $V,$ the piezo stage did not move properly.

For frequency $f$ near the resonance frequency $f_{r}$, the ratio of transmitted power $P_{out}$ to input power $P_{in}$ is~\cite{PaikAPL2010}
\begin{equation}
\frac{P_{out}}{P_{in}}\simeq\displaystyle \left|1-\frac{Q_{L}/Q_{e}e^{i\phi}}{1+i2Q_{L}(f-f_{r})/f_{r}} \right|^2,
\label{eq:PoPi}
\end{equation}
where $Q_{L}$ is the loaded quality factor, $Q_{e}$ is the external quality factor, and $\phi$ accounts for slight asymmetries in the resonance line shape~\cite{PaikAPL2010}.
The internal quality factor $Q_{i}$ is extracted from $1/Q_{L}=1/Q_{i}+1/Q_{e}$.
By fitting Eq.~\ref{eq:PoPi} to the data [see Fig.~\ref{fig:Fig3}(a)], we obtained $Q_{e}$ of about $5\times 10^5$, close to the design value obtained  from  microwave simulations.

Figures~\ref{fig:Fig4}(e) and (f) show the extracted $Q_{i}$ and $Q_{e}$  as a function of tuning frequency for different input power applied to the resonator [taken from the same data used for Fig.~\ref{fig:Fig4}(a)].
We note that -131 dBm corresponds to an average of 11 photons in the resonator.
The increase in $Q_{i}$ with power is consistent with loss due to two-level systems coupled to the resonator~\cite{MartinisPRL2005,ZaeillPRB2008}.
Around $f_{Rb}$, our tunable resonator has $Q_{i}\approx$ 35,000 at low power; however, the quality factor appears to decrease as the tuning pin approaches the resonator (higher frequencies).
This apparent reduction in  $Q_{i}$ is actually broadening of the resonance due to mechanical vibration of the Al pin by about 0.7 $\mu m$ during the 160 s interval needed to measure a resonance curve.
At close separation from the resonator, the vibrations of the pin were observable as a roughly few kHz oscillation in $S_{21}$ and also contribute to the apparent decrease of $Q_{e}$.
In subsequent tests on a resonator made from a 220 nm thick aluminum film, we eliminated the vibrations and oscillations in $S_{21}$ by adjusting the air-suspension vibration-isolation system that the refrigerator is mounted on.
Consistent with reduced vibrational broadening of the resonance, we found a much larger $Q_{i}$ than for the Nb resonator.
Moreover, as the pin approached close to the Al resonator (higher frequencies), both $Q_{i}$ and $Q_{e}$ (see Fig. 5) dropped less rapidly than was the case for the Nb resonator [compare with Fig.~\ref{fig:Fig4}(e) and Fig.~\ref{fig:Fig4}(f)].
Examination of Fig.~\ref{fig:Fig5} also reveals that at large separations (near 6.82 GHz), there is an initial drop in $Q_{i}$ by about a factor of 2 and a corresponding sharp increase in $Q_{e}$ by about 10~\%.
The mechanism producing these variations in $Q_{i}$ and $Q_{e}$ at relatively large pin separations are unclear.

In conclusion, we have demonstrated \emph{in situ} fine tuning through $f_{Rb}$ of the resonance frequency of a superconducting high-Q thin-film resonator by a superconducting Al pin.
The tuning system shows resolution of a few kHz around $f_{Rb}$ and stable operation of a few ppb per hour at 15 mK.
Our scheme enables a superconducting microwave resonator to be coupled to $^{87}$Rb atoms, which will be optically trapped around a sub-$\mu$m diameter optical fiber positioned above the inductor of the resonator ~\cite{Hoffman2011}.

The authors acknowledge discussions with J. Lawall and A. Rauschenbeutel.
This research was supported by NSF through the Physics Frontier Center at the Joint Quantum Institute.

\clearpage
\begin{figure}
\caption{(a) Optical image of lumped-element microwave resonator coupled to a transmission line and surrounded by ground plane. Gray regions are Nb and black regions are sapphire. (b) Optical image of Al pad added to Nb interdigitated capacitor.  }
\label{fig:Fig1}
\end{figure}

\begin{figure}
\caption{(a) Schematic of Al pin and piezoelectric translation stage tuning system. (b) Schematic of experimental set-up.
Microwaves from port 1 of the vector network analyzer (VNA) are sent to the resonator on the mixing chamber
through microwave lines and cold attenuators.
After the resonator the signal passes through two isolators, is amplified by low noise amplifiers (LNAs) at 3.6 K and 300 K, and travels back to port 2 of the VNA. (c) Simulated resonance frequency $f_{r}$ versus the fraction of a finger in the interdigitated capacitor.}
\label{fig:Fig2}
\end{figure}

\begin{figure}
\caption{(a) Measured ratio of transmitted to input power ($P_{out}/P_{in}$) versus frequency after coarse tuning. (b) Deviation of the resonance frequency $f_{r}$ from $f_{0}= 6.827815$ GHz versus time over 70 hours. Time required to measure each point was about 2 minutes. }
\label{fig:Fig3}
\end{figure}

\begin{figure}
\caption{(a) Shift of measured (red circles) and simulated (blue squares) resonance frequency $f_{r}$ around $^{87}$Rb hyperfine splitting frequency $f_{Rb}$ versus distance between Al pin and resonator. (b) $f_{r}$ - $f_{Rb}$ versus number $N$ of piezo steps at different piezo step voltages. (c) and (d) are detailed views of (b) at step voltages of 36 V and 30 V, respectively. (e) and (f) show extracted internal quality factor $Q_{i}$ and external quality factor $Q_{e}$ versus tuned frequency $f_{r}$ for different input power $P_{in}$, respectively. Black dashed line shows the $^{87}$Rb hyperfine splitting frequency $f_{Rb}$. }
\label{fig:Fig4}
\end{figure}

\begin{figure}
\caption{(a) Internal quality factor $Q_{i}$ and (b) external quality factor $Q_{e}$ of the second resonator plotted versus tuned frequency $f_{r}$ for different input power $P_{in}$. Black dashed line shows the $^{87}$Rb hyperfine splitting frequency $f_{Rb}$. The input power to this Al resonator was about -100.5 dBm, which corresponds to about 60,000 photons. }
\label{fig:Fig5}
\end{figure}

\clearpage
\begin{figure}[t]
\centering \includegraphics[width=1\textwidth,angle=0]{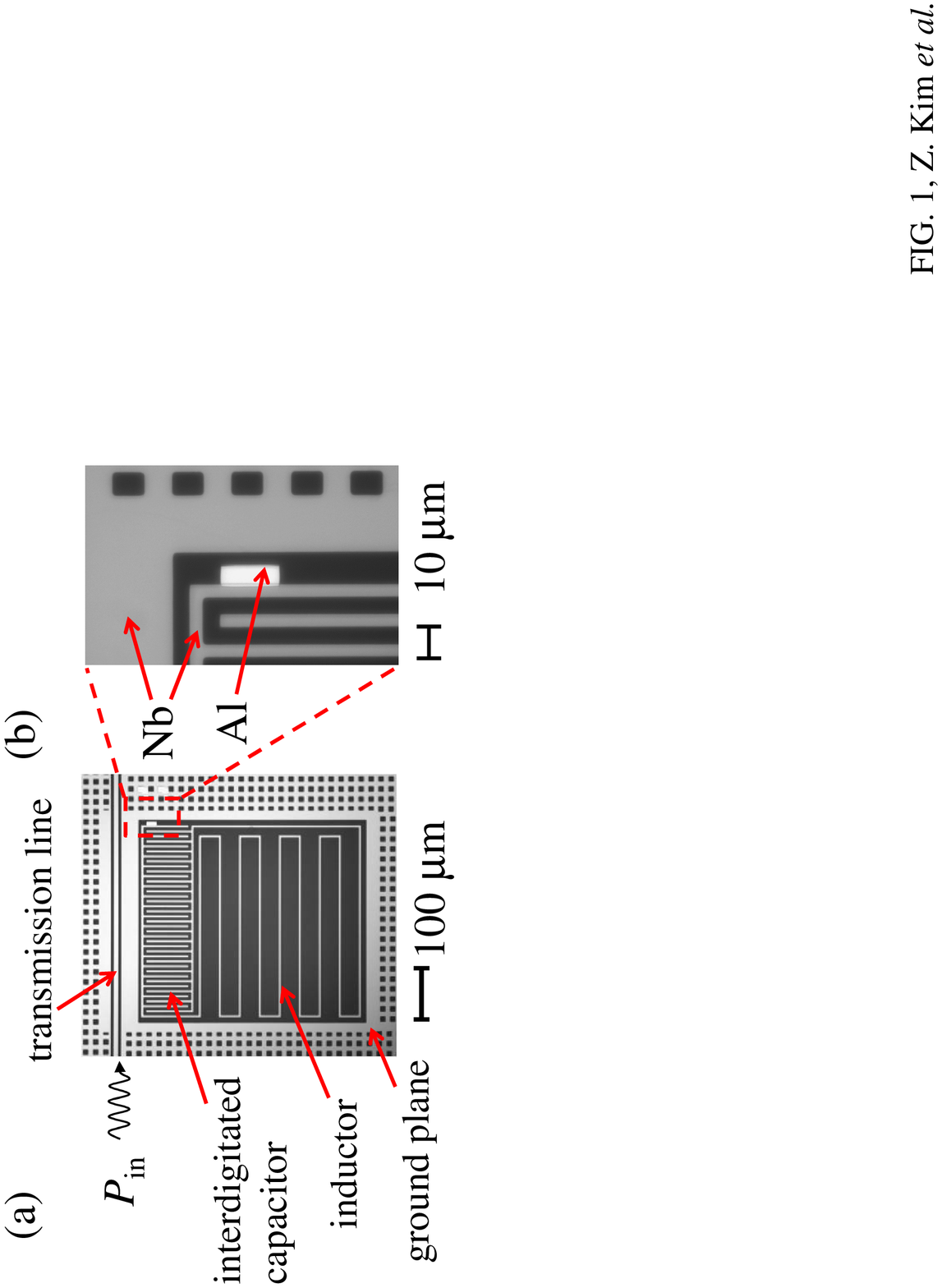}
\end{figure}

\clearpage
\begin{figure}[t]
\centering \includegraphics[width=1\textwidth,angle=0]{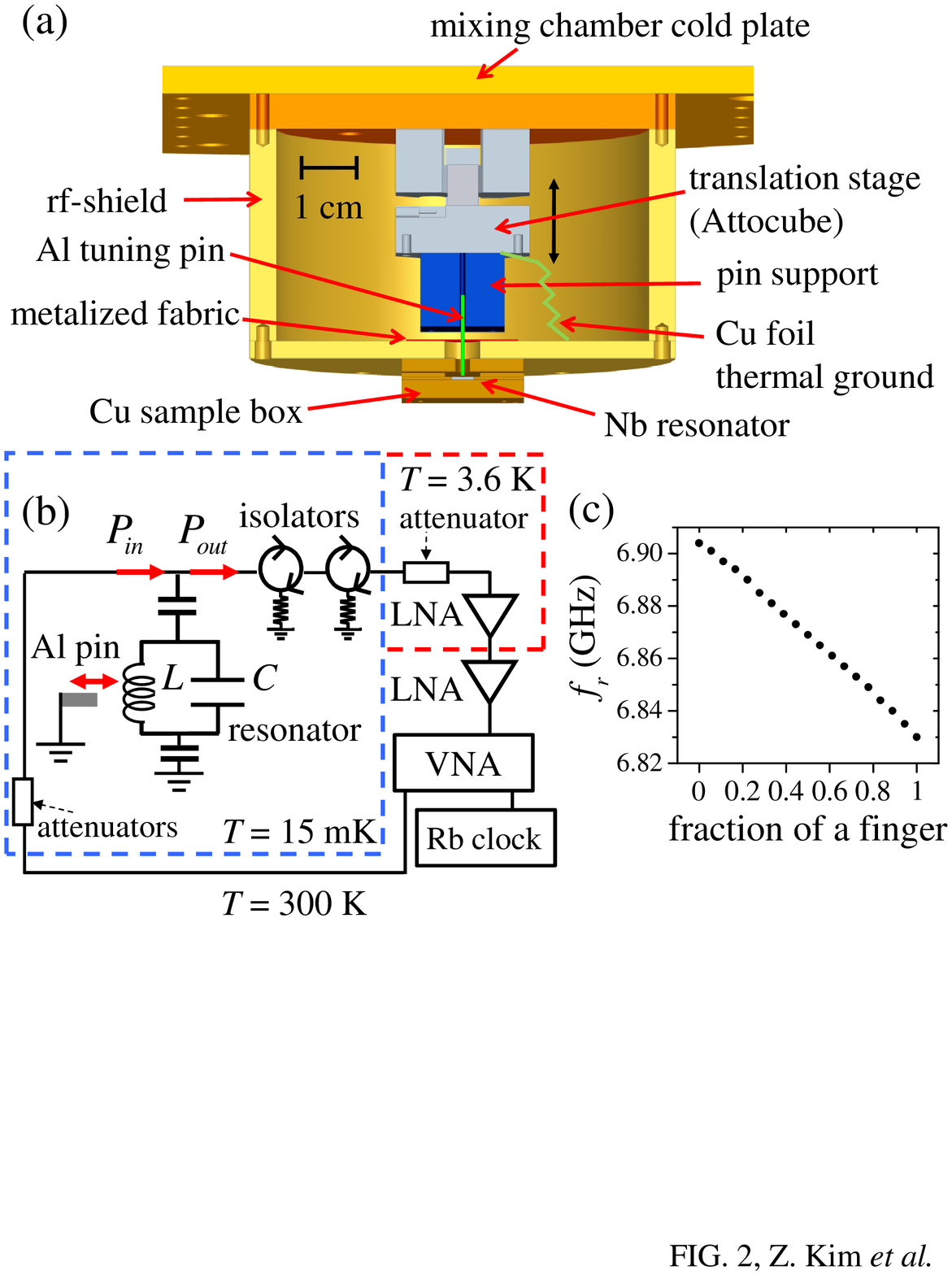}
\end{figure}

\clearpage
\begin{figure}[t]
\centering \includegraphics[width=1.\textwidth,angle=0]{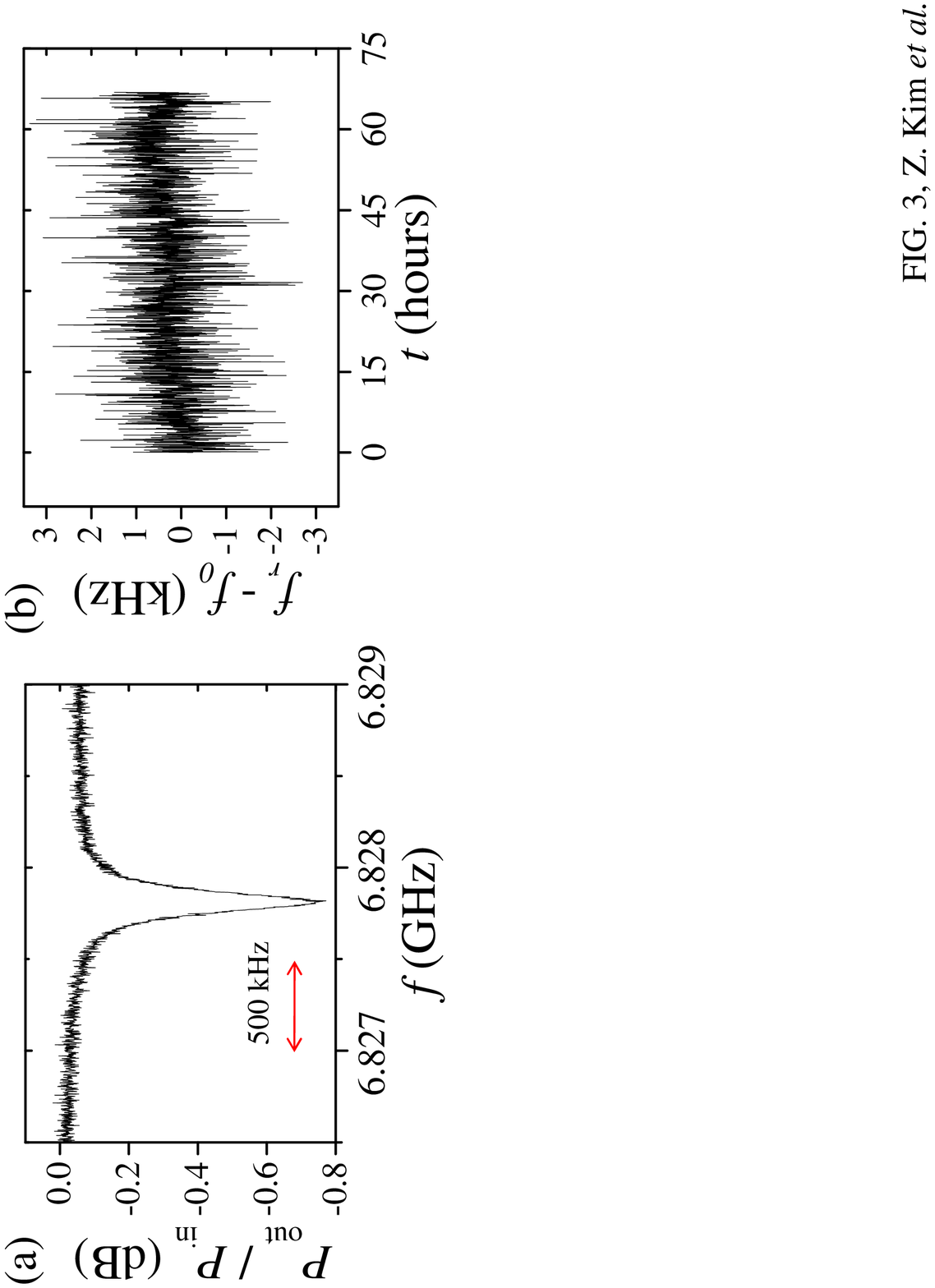}
\end{figure}

\clearpage
\begin{figure}[t]
\centering \includegraphics[width=1\textwidth,angle=0]{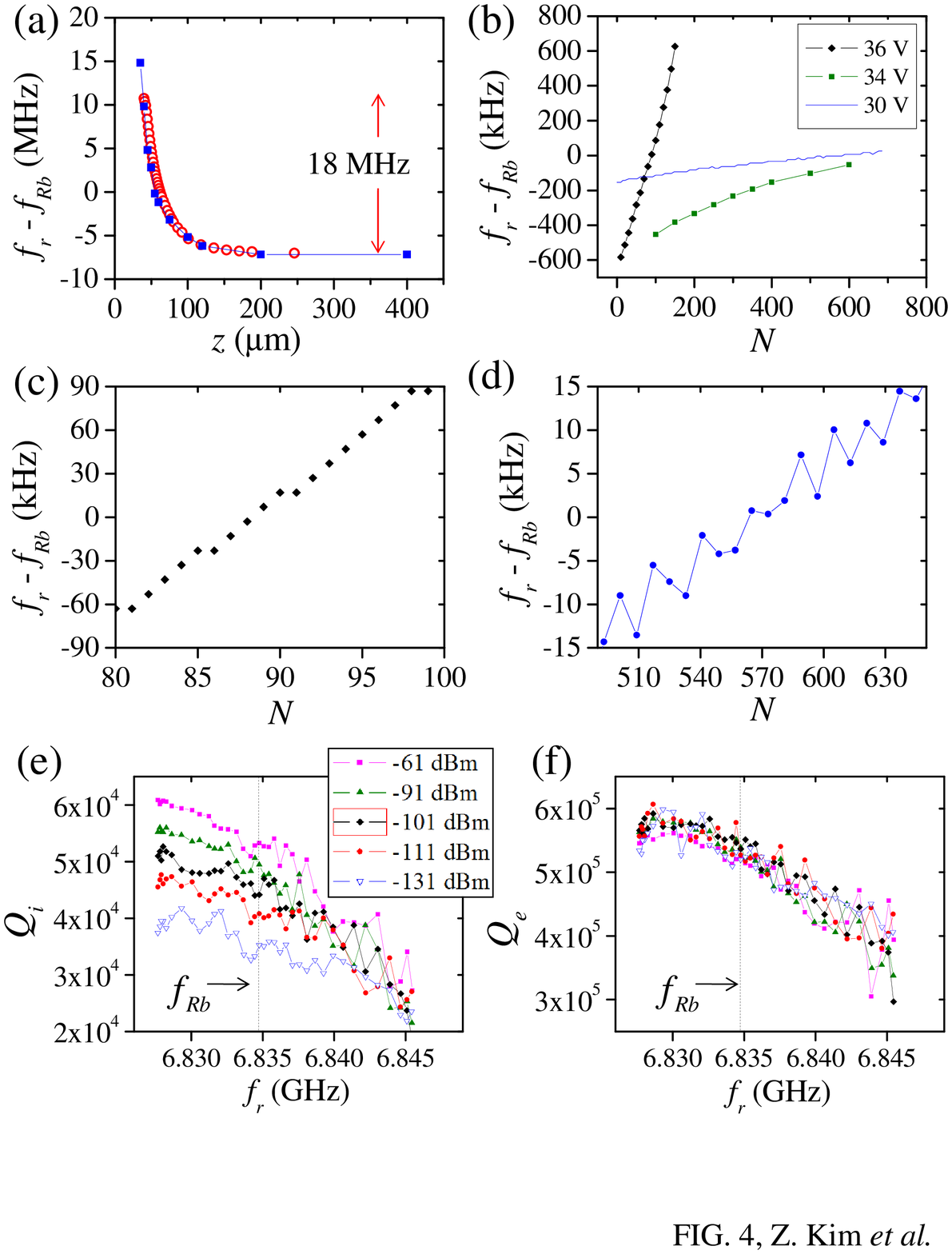}
\end{figure}

\clearpage
\begin{figure}[t]
\centering \includegraphics[width=1\textwidth,angle=0]{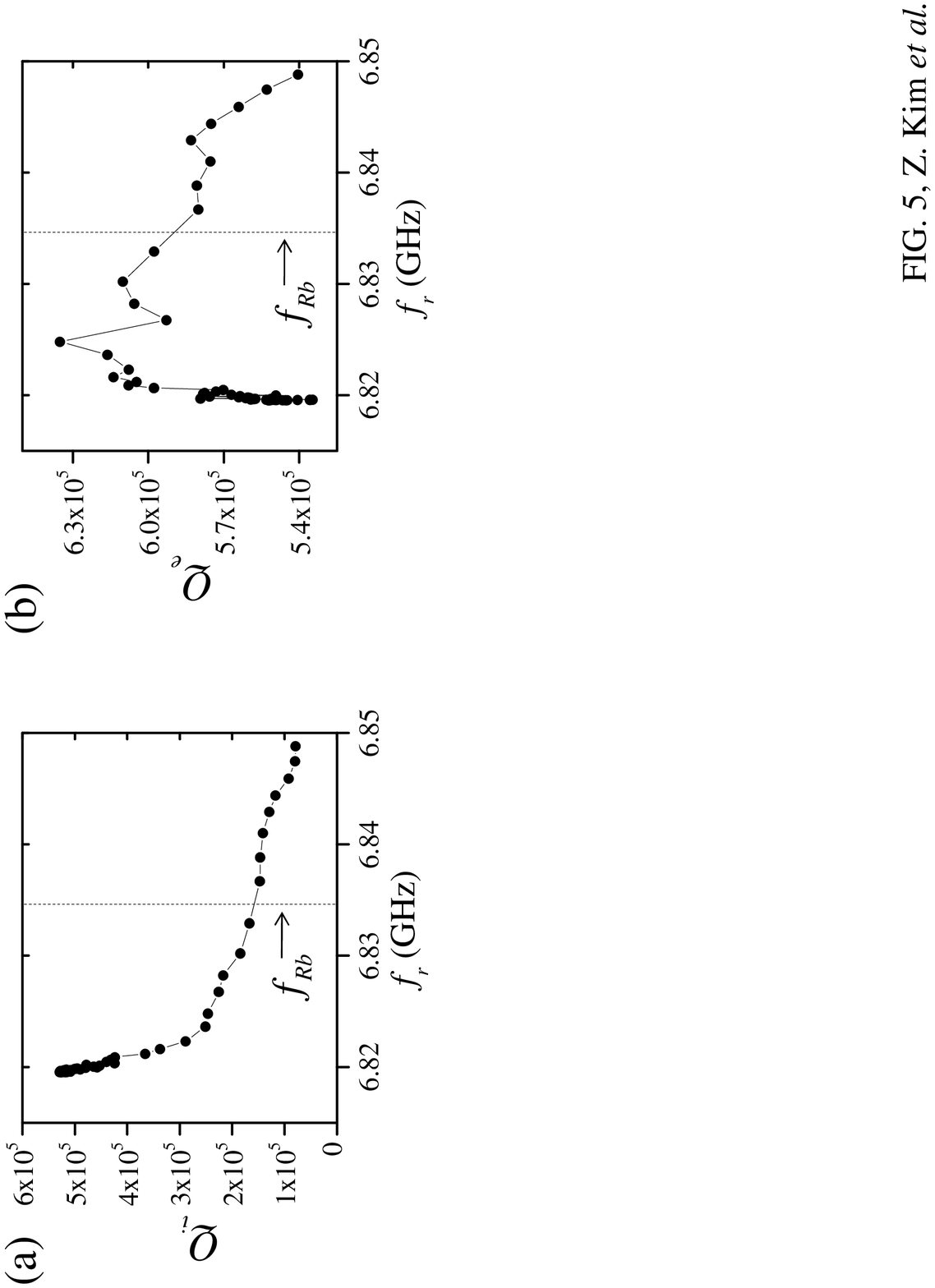}
\end{figure}

\clearpage

\end{document}